\documentclass[10pt,twocolumn,twoside]{article}

\usepackage[T1]{fontenc}
\usepackage{amsmath}
\usepackage{amsfonts}
\usepackage{amssymb}
\usepackage{graphicx}
\usepackage{authblk}
\usepackage[left=2cm,right=2cm,top=2cm,bottom=2cm]{geometry}

\def\onehalf{{\textstyle \frac12}}
\def\ii{{\rm i}}
\def\ssr#1{{\scriptscriptstyle\rm #1}}

\def\jour#1#2#3#4{{\it #1{}} {\bf #2}, #3 (#4)}
\def\lab#1{\label{eq:#1}}
\def\rf#1{(\ref{eq:#1})}
\def\Lie#1{\hbox{\sf #1}}

\def\tsty#1#2{{\textstyle\frac{#1}{#2}}}

\newcommand{\be}{\begin{equation}}
\newcommand{\ee}{\end{equation}}
\newcommand{\bea}{\begin{eqnarray}}
\newcommand{\eea}{\end{eqnarray}}

\title{Unitary rotation and gyration\\
	of pixellated images on rectangular screens}

\author[1]{Alejandro R.\ Urz\'ua}
\author[2,*]{Kurt Bernardo Wolf}
\affil[1]{Posgrado en Ciencias F\'{\i}sicas\\
Universidad Nacional Aut\'onoma de M\'exico}
\affil[2]{Instituto de Ciencias F\'{\i}sicas\\
Universidad Nacional Aut\'onoma de M\'exico\\Av.\ Universidad s/n, 
Cuernavaca, Morelos 62251, M\'exico}
\affil[*]{Corresponding author: bwolf@fis.unam.mx}
\begin{document}
\maketitle

\begin{abstract}
In the two space dimensions of screens in optical systems, 
rotations, gyrations, and fractional Fourier transformations 
form the {\it Fourier\/} subgroup of the symplectic group of 
linear canonical transformations:
$\Lie{U($2$)}_\ssr{F}\subset\Lie{Sp($4$,R)}$. Here we 
study the action of this Fourier group on pixellated images 
within generic rectangular $N_x\times N_y$ screens; its
elements here compose properly and act unitarily, i.e., without 
loss of information.
\end{abstract}

%\ifthenelse{\boolean{shortarticle}}{\ifthenelse{\boolean{singlecolumn}}{\abscontentformatted}{\abscontent}}{}

%-----------------------------------------

\section{Introduction}					\label{sec:one}

Paraxial geometric, wave, and finite optical models with 
two-dimensional plane screens, are covariant with the 
{\it Fourier\/} group $\Lie{U($2$)}_\ssr{F}$. This group 
consists of joint \Lie{SO($2$)} phase space rotations between the 
coordinates ${\bf q}=(q_x,q_y)$ and between their canonically 
conjugate momenta ${\bf p} = (p_x,p_y)$; also it contains joint 
\Lie{SO($2$)} gyrations in the $(q_x,p_y)$ and $(q_y,p_x)$ planes; 
and finally, of $\Lie{SO($2$)}_x\otimes\Lie{SO($2$)}_y$
fractional Fourier transformations that rotate independently
the $(q_x,p_x)$ and $(q_y,p_y)$ planes, and all their compositions. 
In the geometric model, the group $\Lie{U($2$)}_\ssr{F}$ is 
represented by $4\times4$ matrices that are both orthogonal 
and symplectic \cite{SW-00}; in the wave model of images on the 
screen, $f(q_x,q_y)$, ${\bf q}\in{\sf R}^2$, these are subject to 
integral linear canonical transforms \cite{Collins,Moshinsky-Quesne} 
that represent this same group.
In the {\it finite\/} model of optics, where 
images are matrices of values $f(q_x,q_y)$, the 
coordinates $q_x,\,q_y$ are integers that count the 
$N_x\times N_y$ pixels in a rectangular screen, so  
the Fourier group will be represented by square 
$N_xN_y\times N_xN_y$ matrices that are unitary. Of course,
being elements of a group, these $\Lie{U($2$)}_\ssr{F}$ 
transformations can be concatenated and inverted using their simplest 
$4\times4$ representation.

Previously we have considered the action of the $\Lie{U($2$)}_\ssr{F}$ 
Fourier group on finite systems, where the screens were $N{\times}N$
{\it squares\/} \cite{APVW-01,Vic-Rot,KBW-Alieva}. The extension 
to {\it rectangular\/} screens where $N_x\neq N_y$, is not trivial 
because rotations and gyrations to ``angular momentum'' Laguerre-type 
modes require an extended form of symmetry importation 
\cite{Barker0,Barker1}. 

In Sects.\ \ref{sec:two} and \ref{sec:three} we recall the foundations
of the finite model of pixellated optics and the definition of the 
Fourier group in the paraxial geometric model. The fractional Fourier
transforms have their corresponding matrix {\it Fourier-Kravchuk\/} 
transform \cite{AW-97} within the finite model, i.e., they are 
{\it domestic\/} to it. Rotations and gyrations however, require 
{\it importation\/} from the geometric model; this is done
in Sect.\ \ref{sec:four}, where we provide computed examples of
these transformations and show the finite rectangular analogues of
the Laguerre-Gauss modes of wave optics \cite{Ang-Mom-Optics}. 
In Sect.\ \ref{sec:five} we offer some concluding remarks on applications 
to image processing.

%--------------

\section{Continuous and finite oscillator systems}
							\label{sec:two}

The linear finite oscillator system arises as the algebra and group
deformation of the well-known quantum harmonic oscillator,
upon which the continuous position and momentum
coordinates become discrete and finite. 

Let  $\bar Q,\, \bar P$ and $\bar H:= \onehalf(\bar P^2+\bar Q^2-{\it1})$
be the Poisson-bracket or the Schr\"odinger operators of position, momentum and
mode number (do not confuse with the Hamiltonian, which is 
$\bar H+\onehalf{\it1})$, indicating by the over-bar that they refer to the continuous 
model. On the other hand, consider the three components of quantum
angular momentum, designated by the letters $Q\equiv J_1$, $P\equiv -J_2$
and $K\equiv J_3$, and compare their well-known commutation relations
that characterize the oscillator and spin algebras, 
\be
	\begin{array}{rlll}
	\Lie{osc}_1 : & [\bar H,\bar Q]=-\ii\bar P, & [\bar H,\bar P]=+\ii\bar Q,
				& [\bar Q,\par P]=\ii\eta{\it1};\\
	\Lie{su($2$)} : & [K, Q]=-\ii P, & [K,P]=+\ii Q,
				& [ Q, P]=-\ii K.
	\end{array}  \lab{compar}
\ee
The first two commutators in each line are the algebraic form of the 
geometric and dynamical Hamilton equations for the harmonic oscillator 
in phase space, under evolution by $\bar H$ and $K$ respectively. 
The last two commutators however differ, and distinguish between the 
continuous and the finite 
models: in their unitary irreducible representations, the $\Lie{osc}_1$ 
spectrum of $\bar Q$ and $\bar P$ is continuous and fills the real line 
$\sf R$ while that of $\bar H$ is the equally-spaced set $\eta n$ 
($\eta$ fixed) with $n|_0^\infty$ integer; the spectrum of the three 
\Lie{su($2$)} generators on the other hand, in the representation $j$ 
(positive integer or half-integer determined by the eigenvalue $j(j{+}1)$ 
of the  Casimir invariant $\vec J\cdot\vec J\,$), 
is the unit-spaced set $m|_{-j}^j$. This leads us to understand 
$K+j{\it1}$ as the mode number operator of a discrete oscillator 
system that has $2j+1$ modes $n|_0^{2j}$. 

The oscillator Lie algebra $\Lie{osc}_1$ of generators 
${\it1},\,\bar Q,\,\bar P,\,\bar H$
is the {\it contraction\/} of the algebra 
$\Lie{u($2$)}=\Lie{u($1$)}\oplus\Lie{su($2$)}$
with generators ${\it1},\,Q,\,P,\,K$, when we let $j\to\infty$ as the 
number and density of discrete points grows without bound \cite{APW-03}.
This \Lie{u($2$)} can be called the {\it mother\/} algebra of 
the finite oscillator model. The wavefunctions in each model
are the overlaps between the eigenfunctions of their mode generator
and their position generator; they are the Hermite-Gauss (HG) functions
$\Psi_n^\ssr{HG}(q)$ in the continuous model, and 
{\it Kravchuk functions\/} on the
discrete position points of the finite model, given by quantum 
angular momentum theory as Wigner {\it little-d\/} functions 
\cite{Bied-Louck,Atak-Suslov}, for the angle $\onehalf\pi$ between 
$J_1$ and $J_3$, 
\be   \begin{array}{rcl}
	 \Psi_n^{(j)}(q) &:=& d^j_{n-j,q}(\onehalf\pi)\\[3pt] 
	 &=& \displaystyle
		\frac{(-1)^n}{2^j} \sqrt{\Big({2j\atop n}\Big)
				\Big({2j\atop j{+}q}\Big)}\,
				K_n(j{+}q;\,\onehalf,2j),\\[10pt]
	K_n(s;\onehalf,2j)&=&{}_2F_1(-n,-s;-2j;2)=K_s(n;\onehalf,2j),
				\end{array}			\lab{Kravpol}
\ee
where $s|_0^{2j}$, $n|_0^{2j}$, $q|_{-j}^j$, $K_n(s;\onehalf,2j)$ is a symmetric 
Kravchuk polynomial \cite{Kravchuk}, and ${}_2F_1(a,b;c;z)$ 
is the Gauss hypergeometric function. These functions form 
{\it multiplets\/} under \Lie{su($2$)} that have been detailed 
in several papers \cite{AW-97,Echaya}, where they are shown 
to possess the desirable properties of the continuous $HG$ modes. 
For the lowest $n$'s, the points $\Psi_n^{(j)}(q)$ fall closely on the
continuous $\Psi_n^\ssr{HG}(q)$, while for higher $n$'s, they
alternate in sign between every pair of neighbour points,
\be
	\Psi_{2j-n}^{(j)}(q) = (-1)^q \Psi_n^{(j)}(q).
			\lab{sign-change}
\ee

%---------------------------

\section{The Fourier algebra and group}
				\label{sec:three}
				
Consider now two space dimensions $k\in\{x,y\}$, the two
momentum operators, the corresponding two independent 
mode operators, and the single {\it1} (with $\eta_k=1$). 
The $\Lie{osc}_2$ Lie algebra thus generalizes \rf{compar} to
\be
	\begin{array}{c}
	[\bar H_k,\bar Q_{k'}]=-\ii\delta_{k,k'} \bar P_k, \quad
	[\bar H_k,\bar P_{k'}]=\ii\delta_{k,k'} \bar Q_k,\\[5pt]
	\displaystyle [\bar Q_k,\bar P_{k'}]= \ii\delta_{k,k'}{\it 1},
	\end{array} 			\lab{Heisenberg-comm-rel}
\ee
plus $[\bar Q_k,\bar Q_{k'}]=0$ and $[\bar P_k,\bar P_{k'}]=0$.

Out of all quadratic products of $\bar Q_k$ and $\bar P_{k'}$,
one obtains the 10 generators of the symplectic real Lie 
algebra \Lie{sp($4$,R)} of paraxial optics, whose maximal
compact subalgebra is the {\it Fourier\/} algebra 
$\Lie{u($2$)}_\ssr{F}$ \cite{SW-00}. This algebra 
contains four up-to second degree differential operators 
that we identify as the generators of Fourier transformations 
(FT's) and other phase space rotations,  
\bea
 \hbox{symmetric FT }  \bar L_0 &:=&\tsty14(\bar P_x^2{+}
				\bar P_y^2{+}\bar Q_x^2{+}\bar Q_y^2{-}2{\it1})\nonumber\\
					&=&\onehalf(\bar H_x {+} \bar H_y),
								\lab{isotropic-FT}\\
 {}\!\!\!\!\!\!\hbox{antisymmetric FT }   \bar L_1 &:=& \tsty14(\bar P_x^2{-}
				\bar P_y^2{+}\bar Q_x^2{-}\bar Q_y^2)\nonumber\\
					&=&\onehalf(\bar H_x {-} \bar H_y),
					\lab{anti-FT}\\
 \hbox{gyration }   \bar L_2 &:=& \tsty12(\bar P_x\bar P_y
		+\bar Q_x \bar Q_y),  \lab{gyrator-gen}\\
 \hbox{rotation }   \bar L_3 &:=& \tsty12(\bar Q_x \bar P_y
			-\bar Q_y \bar P_x)=:\onehalf\bar M, \lab{ang-mom}
\eea
where $\bar M= 2\bar L_3$ is the physical angular momentum operator.
Their commutation relations are
\be
 [\bar L_0, \bar L_{\ell}]= 0, \quad [ \bar L_i,\,\bar L_j]= \ii\,\bar L_\ell,
				\lab{u2-comm-rel}
\ee
where the indices  $i,\,j,\,\ell $ are a cyclic permutation of  
$1,\,2,\,3 $. Abstractly, \rf{anti-FT}--\rf{ang-mom} 
generate rotations of a 2-sphere.

\begin{figure}
\centering
%\rectangulo{Romboide de modos cartesianos}  
\includegraphics[width=0.8\columnwidth]{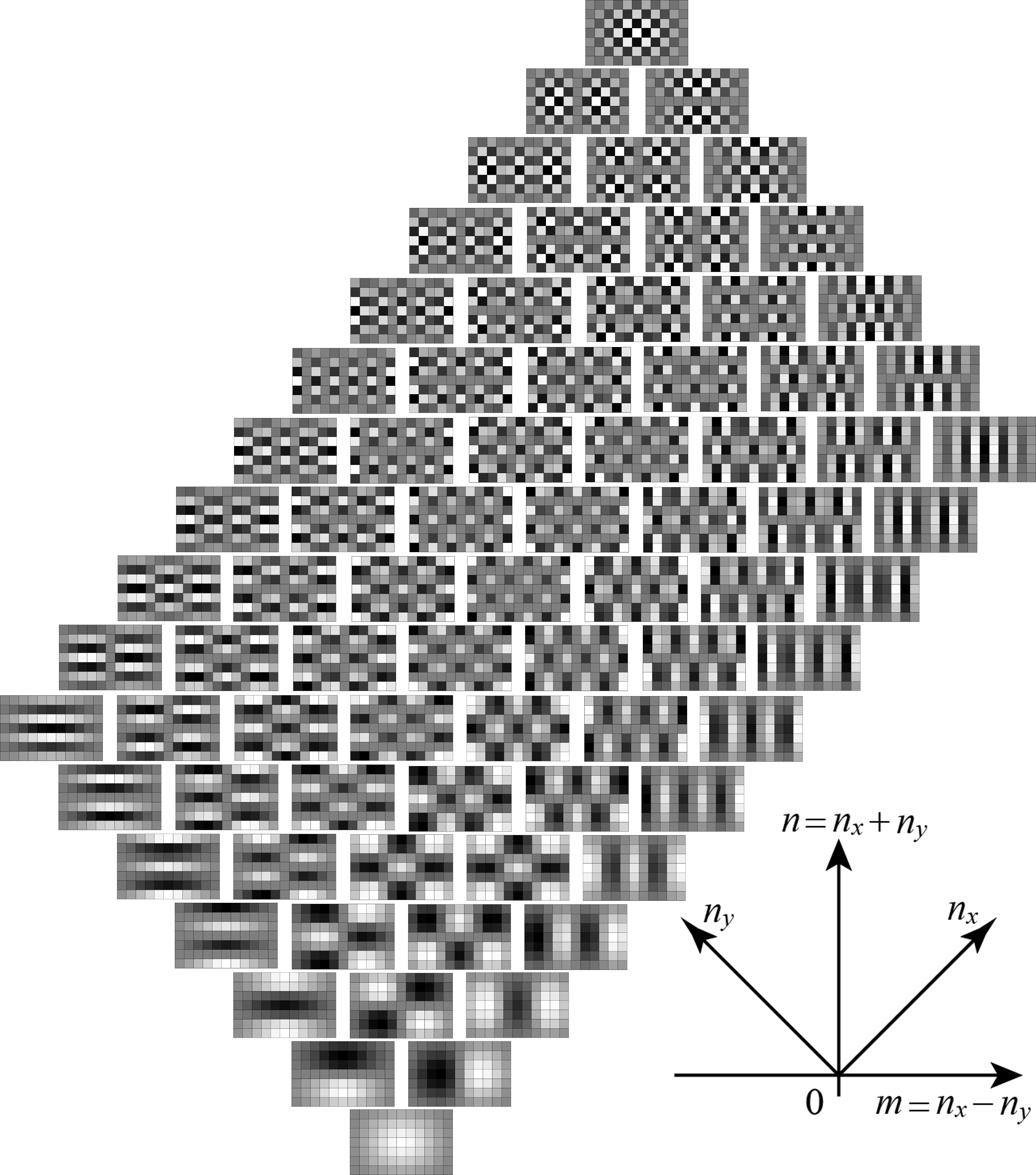}
\caption{Rhomboid with the $11\times7$ Cartesian 
modes $\Psi_{n_x,n_y}^{(5,3)}(q_x,q_y)$ in 
\rf{Carte-Phi}, arranged by $n_x|_0^{10},\,n_y|_0^6$ and
also referred to the axes of {\it total mode\/}  $n=n_x+n_y$ 
and `angular momentum' $m=n_x-n_y$.
In each mode, the pixels are numerated by $q_x|_{-5}^5,\,q_y|_{-3}^3$,
where the $(q_x,q_y)=(5,3)$ pixels are in the upper-right corners.
The range of grey-level densities, from black to white, is $(-1,1)$} 
\label{fig:modos-cartesianos}
\end{figure}

For the finite oscillator model in two dimensions we consider
the direct sum of two $\Lie{su($2$)}$ algebras, whose generators 
form a vector basis for  ${\Lie{su($2$)}}_x\oplus{\Lie{su($2$)}}_y$, 
which is (accidentally) homomorphic to 
the four-dimensional rotation algebra \Lie{so($4$)} \cite{Bied-Louck}.
We choose the representation of this algebra to be $(j_x,j_y)$, 
determined by the values of the two independent Casimir operators
in ${\Lie{su($2$)}}_x\oplus{\Lie{su($2$)}}_y$.
The spectra of positions in the $x$- and $y$-directions will 
thus be $q_k|_{-j_k}^{j_k}$, as will the corresponding spectra 
of momenta, and modes are numbered by $n_k|_0^{2j_k}$. We
interpret the positions as the coordinates of pixels or points in an 
$N_x\times N_y = (2j_x{+}1)\times(2j_y{+}1)$ rectangular array. 
The two-dimensional finite harmonic oscillator 
functions are real and the Cartesian products of Kravchuk functions
\rf{Kravpol} in the two coordinates are \cite{APVW-01},
\be 
	\begin{array}{r}
 	 \Psi_{n_x,n_y}^{(j_x,j_y)}(q_x,q_y) :=
	 \Psi_{n_x}^{(j_x)}(q_x)\Psi_{n_y}^{(j_y)}(q_y),\\
		q_x|_{-j_x}^{j_x},\ n_x|_{0}^{2j_x},\  
		q_y|_{-j_y}^{j_y},\ n_y|_{0}^{2j_y}.
			\end{array}   \lab{Carte-Phi}
\ee
There are thus $N_xN_y$ two-dimensional Kravchuk functions
that can be arranged along axes of {\it total mode\/} $n:=n_x+n_y$ 
and  $m:=n_x-n_y$ into 
the rhomboid pattern shown in Fig.\ \ref{fig:modos-cartesianos}.
As eigenvectors of commuting operators in the Lie algebra, the
Cartesian modes \rf{Carte-Phi} are orthonormal and complete
under the natural inner product,
\bea 
   {}\!\!\!\!\!\!\!\!\!\sum_{n_x,n_y}\!\!\Psi_{n_x,n_y}^{(j_x,j_y)}(q_x,q_y)^*
		\Psi_{n_x,n_y}^{(j_x,j_y)}(q'_x,q'_y)
			&=&\delta_{q_x,q'_x}\delta_{q_y,q'_y},
			\lab{ortho}\\
	{}\!\!\!\!\!\!\!\!\!\sum_{q_x,q_y}\!\!\Psi_{n_x,n_y}^{(j_x,j_y)}(q_x,q_y)^*
		\Psi_{n'_x,n'_y}^{(j_x,j_y)}(q_x,q_y)
			&=&\delta_{n_x,n'_x}\delta_{n_y,n'_y}.
			\lab{complete}
\eea
We shall assume throughout that $j_x>j_y$; when $j_x=j_y$
there will be evident simplifications.

Among the Cartesian modes $\Psi_{n_x,n_y}^{(j_x,j_y)}(q_x,q_y)$ 
in \rf{Carte-Phi} and Fig.\ \ref{fig:modos-cartesianos} we note
that in the lower triangle, where $0\le n \le 2j_y<2j_x$, the
right and left extremes of each $n={}$constant row exhibit $n_x$
and $n_y$ nodes (changes of sign between pixel neighbours) respectively,
both of which can be accommodated within the pixels of the $N_x$ columns 
and $N_y$ rows. As we enter the middle rhomboid, where 
$2j_y< n < 2j_x$, the right extreme $n_x=n\le N_x-1$ can 
accomodate its vertical nodes within the horizontal length of 
the screen, but its left extreme cannot do so among its $N_y$ rows, 
so only $N_y-1$ nodes become horizontal while the rest must remain 
vertical. Finally in the upper triangle, where 
$2j_x\le n \le 2(j_x{+}j_y)$, there will be both vertical as 
well as horizontal nodes in all modes. From \rf{sign-change} 
it follows that 
\be 
	\Psi_{2j_x-n_x,2j_y-n_y}^{(j_x,j_y)}(q_x,q_y)
		= (-1)^{q_x+q_y}\Psi_{n_x,n_y}^{(j_x,j_y)}(q_x,q_y),
			\lab{two-sign-change}
\ee
so that the upper triangle reproduces (top $\leftrightarrow$ bottom,
left $\leftrightarrow$ right) the modes of the lower triangle, but
superposed with a checkerboard of changes of sign. 

The {\it images\/} on the $N_x\times N_y$ pixellated screen
are value arrays $F(q_x,q_y)$ that can be expanded in terms 
of the set of Cartesian modes \rf{Carte-Phi} as
\be
	\begin{array}{rcl}
	F(q_x,q_y)&=&\displaystyle\sum_{n_x,n_y}
		F_{n_x,n_y}\Psi_{n_x,n_y}^{(j_x,j_y)}(q_x,q_y),\\
	F_{n_x,n_y}&=&\displaystyle\sum_{q_x,q_y}
		F(q_x,q_y)\,\Psi_{n_x,n_y}^{(j_x,j_y)}(q_x,q_y),
		\end{array}	\lab{FandPsis}
\ee
as follows directly from linearity, and the orthonormality 
and completeness of the Kravchuk basis.

%---------------------------------

\section{Importation of symmetry}
					\label{sec:four}

In the continuous case, the two-dimensional harmonic oscillator 
mode functions $\Psi_{n_x}^\ssr{HG}(q_x)\*\Psi_{n_y}^\ssr{HG}(q_y)$
can be arranged in a pattern similar to Fig.\ \ref{fig:modos-cartesianos} 
but without an upper bound, forming an `inverted tower' with the same 
lower apex, out of which the total mode number $n:=n_x+n_y$ can grow
indefinitely. Since $\bar L_0$ commutes with all generators in 
\rf{isotropic-FT}--\rf{ang-mom} of the $\Lie{u($2$)}_\ssr{F}$ 
algebra, functions with the same total mode number $n$ will transform 
among themselves under the whole Fourier $\Lie{U($2$)}_\ssr{F}$ group.
Functions with same total mode number $n$ form \Lie{su($2$)} {\it multiplets\/} 
where the range of $n_x-n_y=:m|_{-n}^{n}$ is spaced by two units.
This is equivalent to have multiplets of spin $\lambda:=\onehalf n$, 
with $\mu:=\onehalf m$ playing the role of angular momentum projection 
on a `3'-axis. 

In the finite model however, the generators of the algebra 
$\Lie{su($2$)}_x\oplus\Lie{su($2$)}_y$ can raise and lower
the modes only along the $n_x$ or $n_y$ directions of Fig.\ 
\ref{fig:modos-cartesianos}, but not {\it horizontally}, i.e., 
from one value of $\mu=\onehalf(n_x-n_y)$ to its neighbours. 
{\it Symmetry importation\/} consists in defining 
linear transformations among the states of the finite system 
using the linear combination coefficients provided by  
continuous models \cite{Barker0,Barker1}.

\subsection{Rotations}

The `physical' angular momentum operator $\bar M$ in 
\rf{ang-mom} generates rotations 
$\bar{\cal R}(\theta):=\exp(-\ii\theta\bar M)$ in
the continuous model; this we now import to the finite
model by simply eliminating the over-bar in the notation. 
So, because $M=2L_3$, ${\cal R}(\theta)$ 
is a rotation around the `3'-axis of the sphere by the 
{\it double\/} angle $2\theta$. Since the eigenvalues of 
$L_0$, $\lambda:=\onehalf n=\onehalf(n_x{+}n_y)$, are 
invariant under ${\cal R}(\theta)$, the $2\lambda+1$ 
eigenstates of $L_1$, characterized by the difference
eigenvalues $\mu:= \onehalf(n_x-n_y)$, will mix with 
linear combination coefficients given by Wigner 
little-$d$ functions $d_{\mu,\mu'}^\lambda(2\theta)$ 
\cite{Vic-Rot,Bied-Louck,Echaya}. (Note that the usual 
1-2-3 numbering of axes is rotated to 2-3-1.)

To act on the Cartesian finite oscillator states 
$\Psi_{n_x,n_y}^{(j_x,j_y)}(q_x,q_y)$ in \rf{Carte-Phi},
we note the shape of the rhomboid in 
Fig.\ \ref{fig:modos-cartesianos}, and define
their rotation (initially as a conjecture) by 
\be
	 {\cal R}(\theta):\Psi_{n_x,n_y}^{(j_x,j_y)}(q_x,q_y) :=
		\!\!\!\!\!\!\sum_{n'_x+n'_y=n}\!\!\!\!\!\!
		d^{\lambda(n)}_{\mu,\mu'}(2\theta)
				\Psi_{n'_x,n'_y}^{(j_x,j_y)}(q_x,q_y),
						\lab{rotated-states}
\ee
where the values of spin $\lambda=\lambda(n)$ 
and their projections $\mu(j_x,j_y;n_x,n_y)$ must now
be examined with some care.
The rhomboid contains three distinct intervals of $n$ that
should agree with the correct angular momentum $\lambda$ 
of all imported \Lie{su($2$)} multiplets in the 
horizontal rows of Fig.\ \ref{fig:modos-cartesianos}.

\begin{figure}
\centering
%\rectangulo{Rotaci\'on de modos cartesianos}  
\includegraphics[width=0.8\columnwidth]{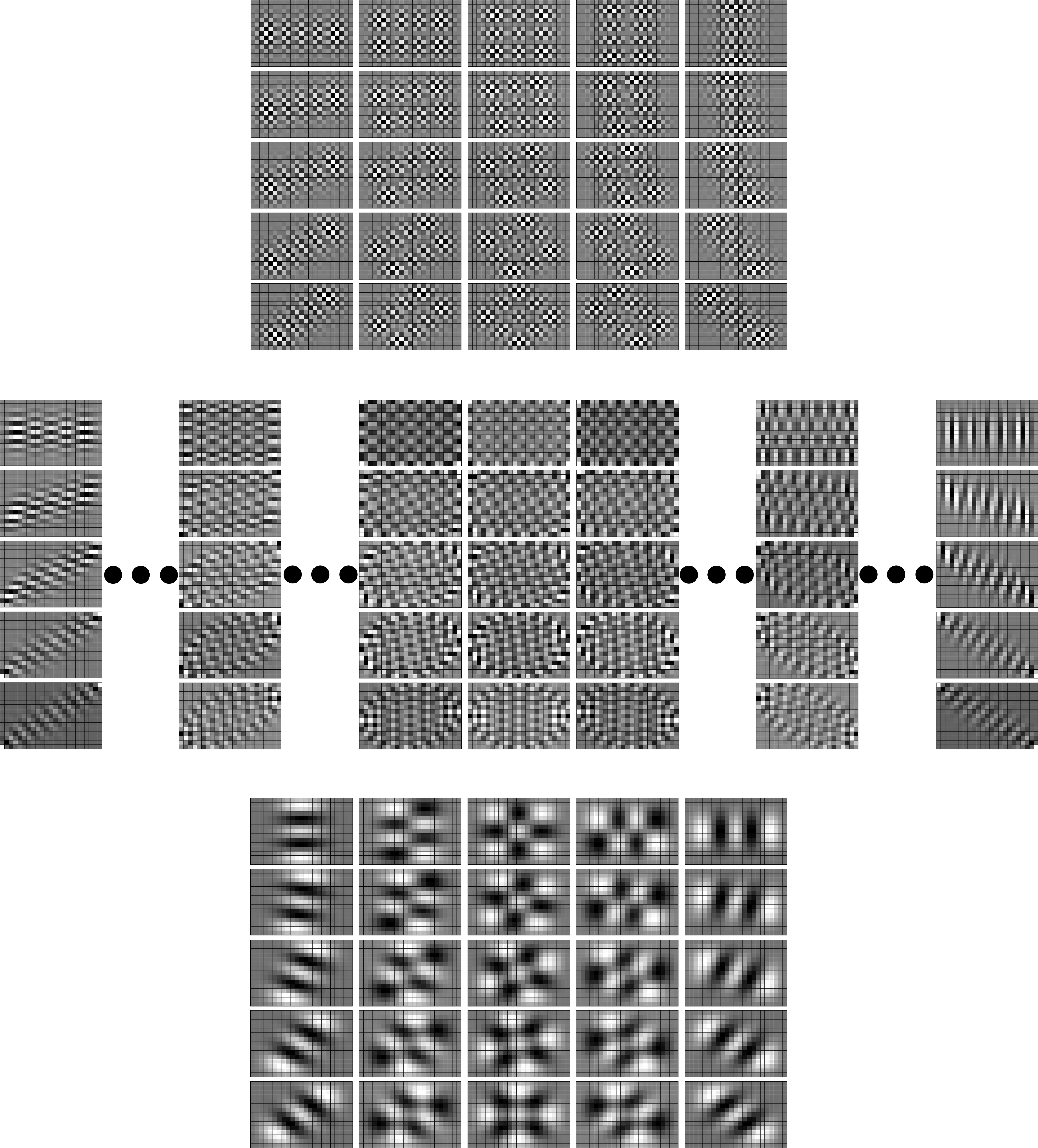}
\caption{Rotations by $\theta=0$, $\frac14\pi$, and 
$\frac12\pi$, of selected Cartesian Kravchuk modes 
$\Psi_{n_x,n_y}^{(11,7)}(q_x,q_y)$ in 
\rf{rotated-states}--\rf{tres-intervalos}. 
Bottom: the 5 states of the level $n=4$ ($\lambda=2$) 
in the lower triangle. Middle: 5 selected states of
the level $n=18$ ($\lambda=7$) in the mid rhomboid. 
Top: the 5 states of the level $n=32$ ($\lambda=2$). 
See that all rotate in the same direction.} 
\label{fig:rota-modos}
\end{figure}

As we have assumed $j_x>j_y$, we recognize that in
each of the three intervals $\lambda(n)$ will be:
\be 
  \begin{array}{lll} 
  	\begin{array}{l}
	\hbox{lower triangle:} \\ 0\le n \le 2j_y,\end{array} & 
	  \left\{\begin{array}{l} \lambda(n)=\onehalf n,\\ 
		 \mu=\onehalf(n_x-n_y),\\ \mu'=\onehalf(n'_x-n'_y), 
		   \end{array}\right.\\[15pt] 
	\begin{array}{l}
	\hbox{mid rhomboid:} \\ 2j_y < n < 2j_x,\end{array} & 
	  \left\{\begin{array}{l} \lambda(n)=j_y,\\ 
		 \mu=j_y-n_y,\\ \mu'=j_y-n'_y, 
		   \end{array}\right.\\[15pt] 
	\begin{array}{l}
	\hbox{upper triangle:} \\ 2j_x\le n \le 2(j_x{+}j_y),\end{array} & 
	  \left\{\begin{array}{l} \lambda(n)=j_x{+}j_y{-}\onehalf n,\\
		 \mu=\onehalf(n_x{-}n_y){-}j_x{+}j_y,\\ \mu'=\onehalf(n'_x{-}n'_y){-}j_x{+}j_y. 
		   \end{array}\right. 
   \end{array}
					\lab{tres-intervalos}
\ee
The first two cases actually overlap for $n=2j_y$ and the second two
cases for $n=2j_x$, which we adjudicate to the triangles
(when $j_x=j_y$ only the two triangles are present \cite{APVW-01}
and overlap for $n=2j$). 
The rotation of various multiplets of two-dimensional 
Kravchuk modes are shown in Figs.\ \ref{fig:rota-modos}. 

\begin{figure}
\centering
%\rectangulo{Rotaci\'on de im\'agen}  
\includegraphics[width=0.8\columnwidth]{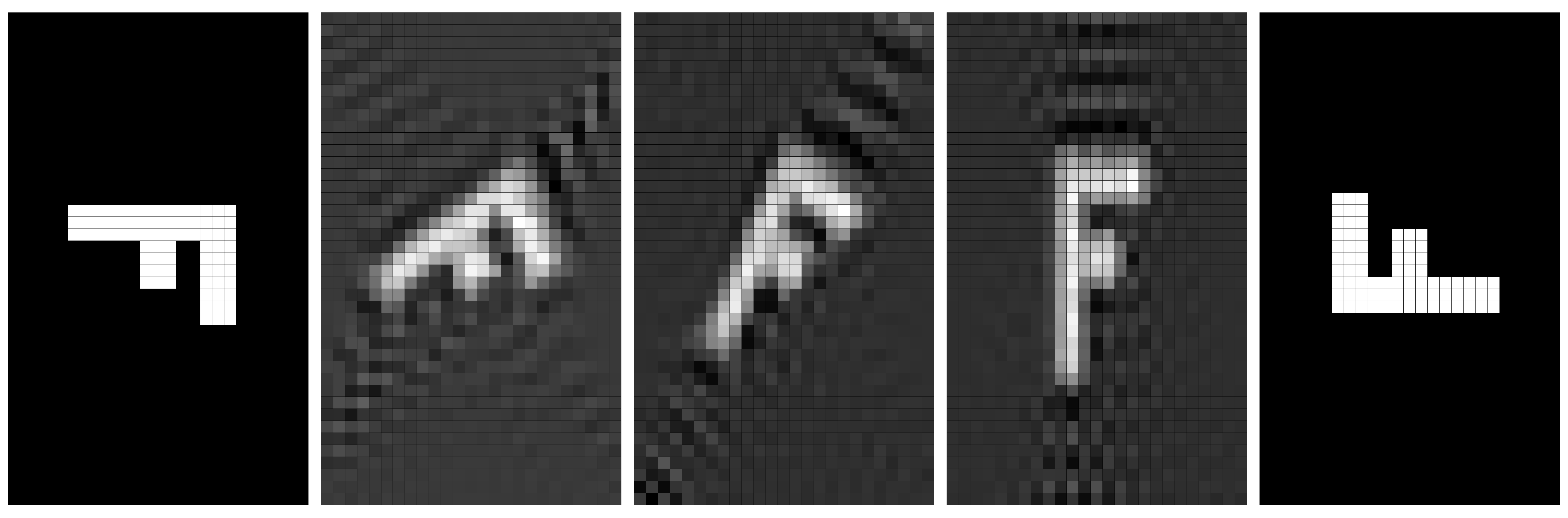}
\caption{On a $41\times25$ pixellated screen, 
$(j_x,j_y)=(20,12)$, successive rotations by 
$\frac16\pi$ of the top image, to $\frac16$, $\frac13\pi$, 
$\frac12\pi$, and (extreme right) $\pi$, of a white image on 
black background. The gray-level scale is adjusted to the
smallest and largest values of the set of pixels; for the 
$\pi=\frac13$ image, these are $(-0.31374,1.38345)$.} 
\label{fig:rota-imagen}
\end{figure}

The rotation of the pixellated {\it images\/} $F(q_x,q_y)$ 
on the $(2j_x{+}1)\times(2j_y{+}1)$ screen follows from 
\rf{FandPsis} and the rotation \rf{rotated-states} 
of the Cartesian basis, 
\be
	\begin{array}{rcl}
	 {\cal R}(\theta)\,{:}\,F(q_x,q_y)\! &=&\displaystyle\!\!\! \sum_{n_x,n_y} \!
	 	F^{(\theta)}_{n_x,n_y}\Psi_{n_x,n_y}^{(j_x,j_y)}(q_x,q_y),\\
	F^{(\theta)}_{n_x,n_y}\!&=&\displaystyle\!\!\!\sum_{q_x,q_y}\!
		F(q_x,q_y)\,{\cal R}(\theta)\,{:}\,\Psi_{n_x,n_y}^{(j_x,j_y)}(q_x,q_y)
		\end{array}	\lab{Rotaim}
\ee
In Fig.\ \ref{fig:rota-imagen} we show the rotation of 
a white on black (1's on 0's) image of the letter ``F''. We note 
the inevitable `Gibbs' oscillations around the sharp
edges of the figure; yet we should stress that the rotated images
were obtained by {\it successive\/} rotations of $\frac16\pi$.
The reconstruction of the original image after six rotations
by $\frac16\pi$ would be impossible with any {\it interpolation\/} 
algorithm applied successively. 

\subsection{Symmetric and antisymmetric Fourier transforms}

In the continuous model, the mode number operators $\bar H_x$ 
and $\bar H_y$ generate 
fractional Fourier transforms \cite{AW-97} through 
$\bar{\cal F}_k(\beta_k) =\exp(-\ii\beta_k\bar H_k)$ that 
multiply the continuous oscillator basis functions 
$\Psi_{n_x}^\ssr{HG}(q_k)$ by phases
$\exp(-\ii n_k\beta_k)$. In the finite oscillator model,
the {\it symmetric\/} fractional Fourier-Kravchuk transform
${\cal K}_\ssr{S}(\chi):=
\exp(-2\ii\chi L_0)$ is generated 
by $L_0=\onehalf(K_x{+}K_y)$ as in
\rf{isotropic-FT}. It acts on the Cartesian modes 
only multiplying them by phases,
\be
	\begin{array}{l}
	 {\cal K}_\ssr{S}(\chi)
	 :\Psi_{n_x,n_y}^{(j_x,j_y)}(q_x,q_y)\\ {\qquad} =
	 			\exp[{-\ii\chi(n_x+n_y)}]\,\Psi_{n_x,n_y}^{(j_x,j_y)}(q_x,q_y),
			\end{array} \lab{Sym-Fouriered-states}
\ee
and commutes with all transformations in the Fourier group.

On the other hand, a rotation by $2\beta$ around the 1-axis
is generated by $L_1 =\onehalf(K_x{-}K_y)$ 
in \rf{anti-FT} to produce the {\it antisymmetric fractional\/} 
Fourier-Kravchuk transforms, ${\cal K}_\ssr{A}(\beta)
:=\exp(-2\ii\beta L_1)$ in the group 
$\Lie{SU($2$)}_x\oplus\Lie{SU($2$)}_y$: 
\be
	\begin{array}{l}
	 {\cal K}_\ssr{A}(\beta)
	 :\Psi_{n_x,n_y}^{(j_x,j_y)}(q_x,q_y)\\ {\qquad} =
		\exp[{-\ii\beta(n_x-n_y)}]\,\Psi_{n_x,n_y}^{(j_x,j_y)}(q_x,q_y).
			\end{array} \lab{Fouriered-states}
\ee
As with rotations, they can be applied to arbitrary images 
using the decomposition in \rf{FandPsis} on the 
pixellated screen. Both ${\cal K}_\ssr{S}(\chi)$
and ${\cal K}_\ssr{A}(\beta)$ are `{\it domestic\/}' within 
${\Lie{SU($2$)}}_x\oplus{\Lie{SU($2$)}}_y$ but they
mesh apropriately with the imported rotations.

\subsection{Gyrations}

\begin{figure}
\centering 
%\rectangulo{Giraci\'on de modos} 
\includegraphics[width=0.8\linewidth]{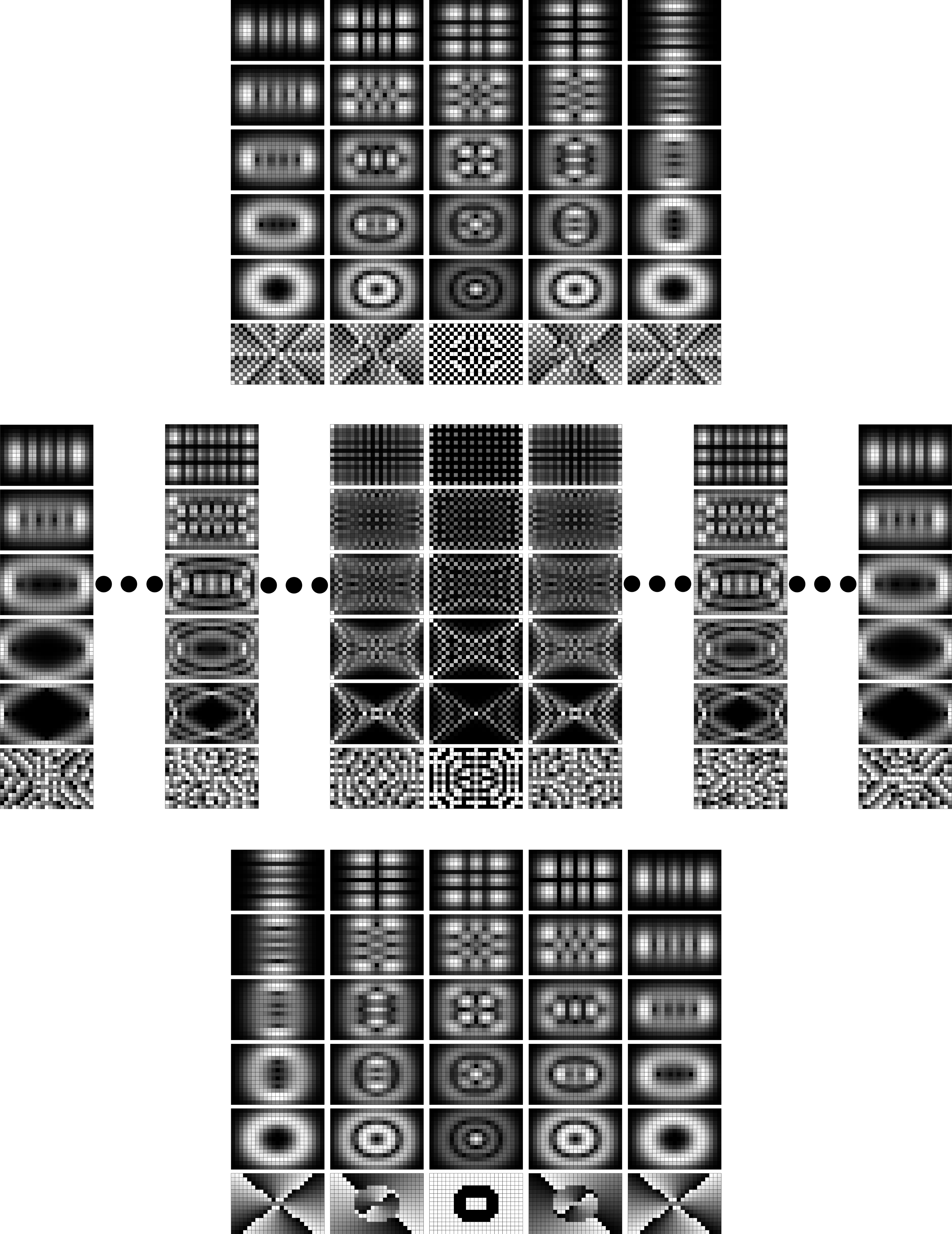}
\caption{Gyrations of selected Cartesian Kravchuk modes 
$\Psi_{n_x,n_y}^{(11,7)}(q_x,q_y)$ in \rf{Gamma--gyrated-states}
by $\gamma=0$, $\frac1{16}\pi$ , and $\frac18\pi$, 
$\frac3{16}\pi$, and $\frac14\pi$. Since the modes are complex,
we show their absolute values; the bottom row of each block 
shows the phase of the $\frac14\pi$ gyration. 
As in Fig.\ \ref{fig:rota-modos}, we display multiplets 
in each of the three blocks according to the three intervals 
in \rf{tres-intervalos}.
Bottom: the level $n=4$ ($\lambda=2$) 
in the lower triangle. Middle: selected states in
the mid rhomboid at level $n=18$ ($\lambda=7$). 
Top: states in the level $n=32$ ($\lambda=2$).} 
\label{fig:gira-modos}
\end{figure}

In the continuous model, {\it gyrations\/} by $\gamma$ 
around the 2-axis are generated by $\bar L_2$ in \rf{gyrator-gen}.
For $\gamma=\frac14\pi$ they transform Hermite-Gauss to
Laguerre-Gauss modes and can be realized with simple
paraxial optical setups \cite{BA-06,AB-07,RAC-07}.
They are rotations that result from a rotation by 
$\onehalf\pi$ around the 3-axis (antisymmetric
fractional Fourier transform by angle $\frac14\pi$), 
a rotation $\gamma$ around the new 1-axis, and back through 
$-\onehalf\pi$ around the new $3$-axis, 
\be
	\bar{\cal G}(\gamma) :=
	\bar{\cal F}_\ssr{\!\! A}(\tsty14\pi)\,
		\bar{\cal R}(\gamma)\,
		\bar{\cal F}_\ssr{\!\! A}(-\tsty14\pi).
									\lab{Gyr-Rot}
\ee
We can thus {\it import\/} gyrations into the finite model 
through replacing $\bar{\cal F}_\ssr{\!\! A} \mapsto
{\cal K}_\ssr{A}$ in \rf{Fouriered-states} and
\rf{Gyr-Rot} \cite{KBW-Alieva}. On the Cartesian Kravchuk modes 
$\Psi_{n_x,n_y}^{(j_x,j_y)}(q_x,q_y)$, gyration will thus act as
\be 
	\begin{array}{l}
	{\cal G}(\gamma):\Psi_{n_x,n_y}^{(j_x,j_y)}(q_x,q_y)\\[3pt]
			\displaystyle 	{\qquad}:=e^{-\ii\pi(n_x-n_y)/4}
			\!\!\!\!\!\sum_{n'_x+n'_y=n} 
			\!\!\!  d^{\lambda(n)}_{\mu,\mu'}(2\gamma)\\
			{\qquad\qquad\qquad}\times
			e^{+\ii\pi(n'_x-n'_y)/4}\,
				\Psi_{n'_x,n'_y}^{(j_x,j_y)}(q_x,q_y).\end{array}
				\lab{Gamma--gyrated-states}
\ee
where $\lambda(n)$, $\mu$ and $\mu'$ are related to $j_x,\,n_x,\,j_y,\,n_y$
through \rf{tres-intervalos}. This set of functions forms also,
as in the Cartesian case, a complete and orthogonal basis for all
images on the pixellated screen. We show the gyration of modes 
$\Psi_{n_x,n_y}^{(j_x,j_y)}(q_x,q_y)$ in Fig.\ \ref{fig:gira-modos} 
for various values of $0\le\gamma\le\frac14\pi$. Note that 
this transformation yields complex arrays of functions, so for
$\gamma=\frac14\pi$ we show their absolute values and phases.

\begin{figure}
\centering
%\rectangulo{Todas las $\Lambda$'s}  
\includegraphics[width=0.7\linewidth]{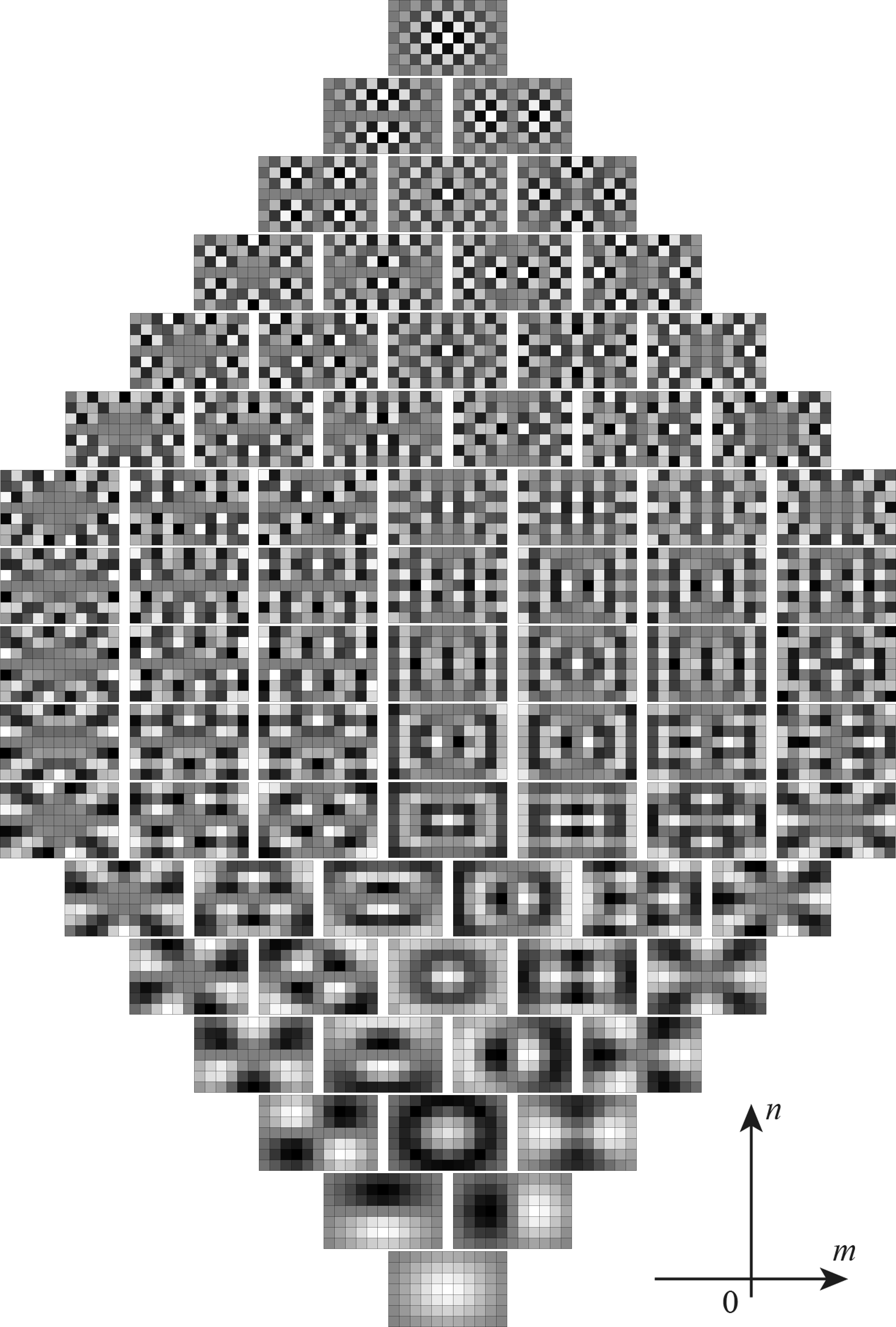}
\caption{The rectangular `Laguerre-Kravchuk' modes $n$
of `angular momentum'$m$,  
$\Lambda_{n,m}^{\!(j_x,j_y)}(q_x,q_y)$ in
\rf{LK-states}. Since the modes are
complex, on the right-hand side $m\ge0$ we show the
density plot of the real part, and on the left-hand
side $m<0$ the imaginary part of the $m>0$ functions.
The $m=0$ modes are real.} 
\label{fig:Todas-Lambdas}
\end{figure}

For $\gamma=\frac14\pi$, \rf{Gamma--gyrated-states} 
defines finite functions characterized by the total mode number 
$n:=n_x{+}n_y$ and an integer `(rectangular) angular 
momentum' number $m=2\mu=n_x{-}n_y$, $|\mu|\le\lambda(n)$ 
constrained 
by \rf{tres-intervalos}, and given by
\be
	\begin{array}{l} \displaystyle
	\Lambda_{n,m}^{\!(j_x,j_y)}(q_x,q_y)
		:=e^{-\ii\pi(n_x-n_y)/4}\!\!\!\!\!\!\sum_{n'_x+n'_y=n} 
			\!\!\!   d^{\lambda(n)}_{\mu,\mu'}(\onehalf\pi)\\
			{\qquad\qquad\quad\qquad}\times e^{+\ii\pi(n'_x-n'_y)/4}\,
				\Psi_{n'_x,n'_y}^{(j_x,j_y)}(q_x,q_y)\\
			{\qquad\qquad\quad\qquad}= \Lambda_{n,-m}^{\!(j_x,j_y)}(q_x,q_y)^*.
				\end{array} \lab{LK-states}
\ee
In the square screen case, when $j_x=j=j_y$, the functions
$\Lambda_{n,m}^{(j)}(q_x,q_y)$ were called Laguerre-Kravchuk
modes \cite{KBW-Alieva}, whose continuous counterparts are 
the well-known Laguerre-Gauss modes. Within rectangular
pixellated screens, the functions \rf{LK-states} are also
orthogonal and complete,
\bea 
	{}\!\!\!\!\!\!\sum_{n,m}\Lambda_{n,m}^{\!(j_x,j_y)}(q_x,q_y)^*\,
		\Lambda_{n,m}^{(j_x,j_y)}(q'_x,q'_y)
			&=&\delta_{q_x,q'_x}\delta_{q_y,q'_y},
			\lab{orthoLambda}\\
	{}\!\!\!\!\!\!\sum_{q_x,q_y}\Lambda_{n,m}^{\!(j_x,j_y)}(q_x,q_y)^*\,
		\Psi_{n',m'}^{(j_x,j_y)}(q_x,q_y)
			&=&\delta_{n,n'}\delta_{m,m'},
			\lab{completeLambda}
\eea
since they are obtained from
the Cartesian states through $\Lie{su($2$)}_\ssr{F}$ unitary
transformations. Although the notion of their angular
momentum is no longer properly valid, they still bear
a recognizable resemblance, as Fig.\ \ref{fig:Todas-Lambdas} 
shows, where the multiplets 
$\{\Lambda_{n,m}^{\!(j_x,j_y)}(q_x,q_y)\}_{m=-\lambda(n)}^{\lambda(n)}$ 
are placed on rows for all total modes $n$ in the three
intervals \rf{tres-intervalos}.

%----------------------------------

\section{Concluding remarks}   \label{sec:five}

In continuous systems, the elements of the 
$\Lie{U($2$)}_\ssr{F}$ Fourier group
can be param\-e\-trized by the angle of the 
central symmetric Fourier transform and three 
$\Lie{SU($2$)}_\ssr{F}$ Euler angles as
\be
	\begin{array}{rcl}
	\bar{\cal D}(\chi;\psi,\theta,\phi)&=&
	   \exp(-\ii\chi \bar L_0)\\		
		&\times&\exp(-\ii\psi \bar L_3)
			\exp(-\ii\theta \bar L_2)
				\exp(-\ii\phi \bar L_3).  
					\end{array} \lab{Euler-angle}
\ee
On the finite $N_x\times N_y$ pixellated screen, 
$\Lie{U($2$)}_\ssr{F}$ is correspondingly realized by 
subgroups of domestic Fourier-Kravchuk
transformations, and imported rotations and gyrations, as
\be 
  \begin{array}{l}
	{\cal D}(\chi;\psi,\theta,\phi) =
	{\cal K}_\ssr{S}(\tsty12\chi)\,
	{\cal K}_\ssr{A}(\tsty12\psi)\,
	{\cal G}(\onehalf\theta)\,
	{\cal K}_\ssr{A}(\tsty12\phi)\\[3pt]
		{\qquad}=
	{\cal K}_\ssr{S}(\tsty12\chi)\,
	{\cal K}_\ssr{A}(\onehalf\psi{+}\tsty14\pi)\,
	{\cal R}(\onehalf\theta)\,
	{\cal K}_\ssr{A}(\onehalf\phi{-}\tsty14\pi).
			\end{array}			\lab{Euler-angle-op}
\ee

The action of the Fourier group is unitary on all  
complex-valued images on $N_x\times N_y$ pixellated screens,
and hence there is no information loss under these
transformations. We must repeat that the algorithm
is not fast, but arguably the slowest, and will 
necessarily involve Gibbs-like oscillations in
pixellated images with sharp `discontinuities'.
As the previous experience with square screens
suggests \cite{Vic-Rot}, smoothing the original values or chopping
the resulting ones can restore the visual fidelity 
of the image, even though unitarity will be lost.
It may be that in experiments where two-dimensional 
beams are sampled at rectangular CCD arrays, bases
of discrete functions are better suited for the
task than their approximation by pointwise-sampled 
Hemite-Gauss oscillator wavefunctions \cite{LEV-KBW}. 

The introduction of rectangular analogues of the 
Laguerre-Gauss states with `angular momentum' is
the direct (but not trivial) generalization of 
those built for square screens in Ref.\ \cite{APVW-01}
and, as there, will predictably allow a unitary map
to screens whose pixels are arranged along polar
coordinates, as done in Ref.\ \cite{APVW-02}. Finally,
it should be noted that all the `discrete' functions,
starting with the Wigner little-$d^j_{n-j,q}(\theta)$, are actually
analytic functions of {\it continuous\/} position $q$ in the
range $-j{-}1<q<j{+}1$, with branch-point zeros at 
$q=\pm(j{+}1)$ and cuts beyond. This property extends
of course to the two-dimensional case for $-j_k{-}1<q_k<j_k{+}1$,
$k\in\{x,y\}$. 
The discrete model can thus also accommodate a continuous 
model of modes in bounded screens, although the unitarity of
the transformations holds only for the discrete integer 
points within.

%----------------------------------

\section*{Acknowledgments}

We thank the support of the Universidad Nacional Aut\'onoma
de M\'exico through the PAPIIT-DGAPA project IN101115
{\it \'Optica Matem\'atica}, and acknowledge the help of
Guillermo Kr\"otzsch with the figures.

\end{document}